\documentstyle[12pt]{article}

\newcommand{\EQ}{\begin{equation}}
\newcommand{\EN}{\end{equation}}
\newcommand{\bea}{\begin{eqnarray}}
\newcommand{\ena}{\end{eqnarray}}
\newcommand{\vs}[1]{\vspace{#1 mm}}
\newcommand{\hs}[1]{\hspace{#1 mm}}

\renewcommand{\d}{\delta}
\newcommand{\e}{\epsilon}
\def\bbox{{\,\lower0.9pt\vbox{\hrule \hbox{\vrule height 0.2 cm
\hskip 0.2 cm \vrule height 0.2 cm}\hrule}\,}}
\newcommand{\dsl}{\pa \kern-0.5em /}

\newcommand{\shalf}{\frac{1}{2}}
\newcommand{\pa}{\partial}
\newcommand{\nn}{\nonumber\\}
\newcommand{\p}[1]{(\ref{#1})}
\newcommand{\lan}{\langle}
\newcommand{\ran}{\rangle}

\begin{document}

\topmargin 0pt
\oddsidemargin 0mm

\renewcommand{\thefootnote}{\fnsymbol{footnote}}
\begin{titlepage}

\setcounter{page}{0}
\begin{flushright}
KL-TH  98/12\\
OU-HET 304 \\
hep-th/9809193
\end{flushright}

\vs{10}
\begin{center}
{\Large\bf AdS$_3$/CFT Correspondence, Poincar\'{e} Vacuum State and 
Greybody Factors in BTZ Black Holes}
\vs{10}

{\large
H.J.W. M\"{u}ller-Kirsten,$^{a,}$\footnote{e-mail address:
mueller1@physik.uni-kl.de}
Nobuyoshi Ohta$^{b,c,}$\footnote{e-mail address: ohta@phys.wani.osaka-u.ac.jp}
and
Jian-Ge Zhou$^{c,}$\footnote{e-mail address: jgzhou@phys.wani.osaka-u.ac.jp}}

\vs{10}
$^a${\em Department of Physics, University of Kaiserslautern, \\
D-67653 Kaiserslautern, Germany} \\
\vs{5}
$^b${\em The Niels Bohr Institute, Blegdamsvej 17, DK-2100 Copenhagen
\O, Denmark}\\

\vs{5}
$^{c}${\em Department of Physics, Osaka University,
Toyonaka, Osaka 560-0043, Japan}

\end{center}
\vs{10}
\centerline{{\bf{Abstract}}}
\vs{5}

The greybody factors in BTZ black holes are evaluated from 2D CFT in the
spirit of AdS$_3$/CFT correspondence. The initial state of black holes in
the usual calculation of greybody factors by effective CFT is described as
Poincar\'{e} vacuum state in 2D CFT. The normalization factor which cannot
be fixed in the effective CFT without appealing to string theory is shown
to be determined by the normalized bulk-to-boundary Green function.
The relation among the greybody factors in different dimensional black
holes is exhibited. Two kinds of $(h,{\bar h})=(1,1)$ operators which
couple with the boundary value of massless scalar field are discussed.

\end{titlepage}

\newpage

\renewcommand{\thefootnote}{\arabic{footnote}}
\setcounter{footnote}{0}

Recently it has been proposed~\cite{M} that string/M-theory on AdS$_d
\times M$ (where $M$ is a proper compact space) is dual to a conformal
field theory (CFT) which lives on the boundary of the anti-de Sitter
(AdS) space. Precise forms of the conjecture~\cite{M} of the AdS/CFT
correspondence have been stated and investigated in refs.~\cite{GKP,W}.
The essence of this conjecture is that supposing the partition functions
of the two theories are equal, the correlation functions in the CFT can
be read off from the bulk theory and vice versa. One of the interesting
examples is the duality between type IIB string theory on AdS$_3 \times
S^3 \times M_4$ and certain 2D CFT~\cite{MS1}-\cite{GKS}, since both
sides of the theories are very well understood. It has been shown that
there is an agreement between the Kaluza-Klein spectrum of supergravity
and the spectrum of certain 2D CFT~\cite{MS1,LB}.

On the other hand, much work on the absorption and Hawking radiation
in 5D and 4D black holes has been done in the semiclassical analysis,
D-brane picture, effective string model and effective 2D
CFT~\cite{DMW1}-\cite{DMW2}. Since 5D and 4D black holes are U-dual
to BTZ black holes~\cite{BTZ}, the entropies of 5D and 4D black holes
can be related to the entropy of the BTZ black holes~\cite{SSH}.
Especially, in the large $N$ limit~\cite{M}, the geometries of 5D
and 4D black holes are AdS$_3 \times M$ (effectively near horizon
region). So one expects to find the explanation of greybody factors
of higher dimensional black holes in the context of BTZ black holes.
In refs.~\cite{BSS,LM}, the greybody factors in BTZ black holes have
been discussed from the semiclassical point of view, where the wave
equations are solved.

In the effective string calculation of greybody factors in 5D and 4D black
holes, the interaction between the scalar fields in the bulk and effective
string is described by~\cite{G}
\bea
S_{int} = \int d^2 x \phi(t,x,{\vec x}=0) {\cal O}(t,x),
\label{int}
\ena
where the integration is over the effective string worldsheet,
${\vec x}$ indicates its location in transverse space, and
${\cal O}(t,x)$ is some local conformal operator in 2D CFT, which
takes the form
\bea
{\cal O}(t,x) = {\cal O}_+ (x^+) {\cal O}_- (x^-),
\label{o}
\ena
where $x^\pm = t \pm x$ and ${\cal O}_+$ and ${\cal O}_-$ are primary fields
of dimension $h_L$ and $h_R$, respectively.

The OPE's of ${\cal O}_+$ and ${\cal O}_-$ with themselves are given by
\bea
{\cal O}_+({\bar z}) {\cal O}_+({\bar w}) &=& \frac{C_{\cal O_+}}
 {({\bar z}-{\bar w})^{2 h_L}} + {\rm less \; singular \; terms}, \nn
{\cal O}_-(z) {\cal O}_-(w) &=& \frac{C_{\cal O_-}}
 {(z - w)^{2 h_R}} + {\rm less \; singular \; terms},
\label{ope}
\ena
where $z= i x^-, {\bar z} = i x^+$ for real $x$ and imaginary $t=-i \tau$.
To compare with the macroscopic decay rate in 5D and 4D black holes,
the initial state is usually thermally averaged, since the black hole
corresponds to a thermal state. This means that one must take finite
temperature two-point correlation functions. Thus the function $G(t,x)$
defined by
\bea
G(t,x) = \lan {\cal O} (-i\tau,x){\cal O}(0,0) \ran_{T_H},
\label{gd}
\ena
is usually chosen as
\bea
G(t,x)= \frac{C_{\cal O_+} C_{\cal O_-}}{i^{2 h_L + 2 h_R}}
 \left( \frac{\pi T_L}{\sinh \pi T_L x^+} \right)^{2 h_L}
 \left( \frac{\pi T_R}{\sinh \pi T_R x^-} \right)^{2 h_R}.
\label{g}
\ena
Since $x^+$ and $x^-$ are linear in ${\bar z}$ and $z$, respectively,
there seems to be some disagreement between eqs.~\p{g} and~\p{ope}. In
other words, it is not clear what kind of initial
state is used to define \p{gd}. Then natural questions arise how to
describe the initial state of black holes in 2D CFT, and how to
explain the greybody factors in 5D and 4D black holes in terms of
those in BTZ black holes. These are the problems we are going to examine.

In order to answer the above questions, we discuss the greybody factors in
BTZ black holes from 2D CFT in the spirit of AdS$_3$/CFT correspondence.
To get explicit description for \p{gd} and \p{g}, we calculate the two-point
correlation functions in the BTZ coordinates by bulk-boundary
correspondence~[1-3,18-21], but here we should include all the coefficients
in the calculation. Using two-point correlation functions in BTZ
coordinates, we evaluate the greybody factors in BTZ black holes, and find
that they agree with the known results even for the coefficients. This
fact gives further evidence to the AdS$_3 \leftrightarrow$ CFT
correspondence. In fact, the calculation heavily relies on Witten's
conjecture~\cite{W}. The result obtained shows that the initial state
of black holes can be described by Poincar\'{e} vacuum state in 2D CFT.
The coordinate transformation between Poincar\'{e} coordinates
$(w_+, w_-)$ and BTZ coordinates $(u_+, u_-)$ induces a mapping of
the operator ${\cal O}(w_+, w_-)$ to ${\cal O}(u_+, u_-)$ by Bogoliubov
transformation, and the operator ${\cal O}(u_+, u_-)$ sees the
Poincar\'{e} vacuum state (which is a natural vacuum state for 2D CFT)
as an excited mixed state.

After explaining the greybody factors in BTZ black holes,
we find that the greybody factors in 5D and 4D black holes can be
described by those in BTZ black holes
in a unified way. This is because in the large $N$ limit~\cite{M}, the
geometries of 5D and 4D black holes turn out to be BTZ $\times S^3 \times M_4$
and BTZ $\times S^2 \times M_5$, respectively, and the two-point correlation
functions in 5D and 4D black holes can be related to those in BTZ black
holes, which are constant multiples of those in BTZ case. The constant
can be determined from the parent 10D supergravity theories~\cite{FMM,T,OZ}.
It is known that the greybody factors in black holes only depend on
the conformal dimension of the operator ${\cal O}(t,x)$, and are
indifferent to the concrete form of the operator. In order to understand
the relation between the physical degrees of freedom in 3D pure gravity
and those of 2D CFT, we next consider an $(h,{\bar h})=(1,1)$ operator
in ${\cal N} =(4,4)$ super CFT (SCFT) based on a resolution of the
orbifold $(T^4)^{Q_1 Q_5}/ S(Q_1 Q_5)$, and the corresponding operator
in 3D gravity obtained by quantizing BTZ black holes in external massless
scalar field. By comparing two operators, we find that the contribution
from different scalars $x_A$ is smeared, and gravity cannot distinguish
between different CFT states with the same expectation value for the operator
${\cal O}(u_+,u_-)$, which shows that gravity is just like thermodynamics
but gauge theory is like statistical mechanics~\cite{M1,M2}.

Now let us consider two-point correlation functions of 2D CFT; in Poincar\'{e}
coordinates many studies of correlation functions in boundary CFT have
been done by the bulk-boundary correspondence~\cite{HSM,FMM,BKL}; in BTZ
coordinates this was analysed in ref.~\cite{K}. To calculate the greybody
factors in BTZ black holes, we need to keep all the coefficients in the
two-point correlation functions, and then compare the greybody factors
extracted from AdS$_3$/CFT correspondence with those obtained from the
semiclassical analysis.

As a first simple exploration, we consider two-point correlation functions
of the operator coupling to the boundary value of the massive scalar field.
In terms of Poincar\'{e} coordinates, the AdS$_3$ metric is
\bea
ds^2 = \frac{l^2}{y^2} ( dy^2 + dw_+ dw_-).
\label{poi}
\ena
For simplicity, we choose $l=1$ in the following discussions.

The Euclidean action of massive scalar field with mass $m$ in AdS$_3$
space is
\bea
S(\phi)= \shalf \int dy dw_+ dw_- \sqrt{g} \left(
g^{\mu\nu} \pa_\mu \phi \pa_\nu \phi + m^2 \phi^2 \right),
\label{euc}
\ena
which has solution behaving as $\phi(y,w_+,w_-) \to y^{2h_-}
\phi_0 (w_+,w_-)$ when $y \to 0$. The boundary value $\phi_0 (w_+,w_-)$
has dimension $2h_-$ which couples to an operator ${\cal O}(w_+,w_-)$
of dimension $2h_+$ with the parameters $h_\pm$ defined by
\EQ
h_\pm = \shalf (1 \pm \sqrt{1+m^2}).
\EN
The normalized bulk-to-boundary Green function in Poincar\'{e} coordinates
is~\cite{FMM}
\bea
K_P (y,w_+,w_-;w_+',w_-') =
\frac{\Gamma(2h_+)}{\pi \Gamma(2h_+ -1)} \left[ \frac{y}{y^2 + (w_+ -w_+')
(w_- -w_-')} \right]^{2h_+},
\label{bbg}
\ena
which has singular behavior when $y \to 0$ as~\cite{W,FMM}
\bea
y^{-2h_-} K_P(y,w_+,w_-;w_+',w_-') \to \d(w_+ -w_+') \d(w_- -w_-').
\label{beh}
\ena
Here we should mention that the behavior \p{beh} determines the
normalization coefficient in \p{bbg}. The solution $\phi(y,w_+,w_-)$ can
be expressed as
\EQ
\phi(y,w_+,w_-) = \int dw_+' dw_-' K_P(y,w_+,w_-;w_+',w_-')\phi_0(w_+',w_-'),
\label{sol}
\EN
where $\phi_0(w_+,w_-)$ is the boundary value of the bulk field
$\phi(y,w_+,w_-)$.

The metric of BTZ black holes is~\cite{BTZ}
\bea
ds^2 = - \frac{(r^2-r_+^2)(r^2-r_-^2)}{r^2} dt^2
 + \frac{r^2}{(r^2-r_+^2)(r^2-r_-^2)} dr^2 + r^2 \left( d\phi
 - \frac{r_+ r_-}{r^2} dt \right)^2,
\label{met}
\ena
with periodic identification $\phi \sim \phi + 2 \pi$, where we have chosen
$l=1$. The mass and angular momentum of BTZ black holes are defined as
\bea
M = r_+^2 + r_-^2, \;\;
J = 2 r_+ r_-.
\ena
It can be shown that the metric of BTZ black holes can be transformed
to that of AdS$_3$ locally by~\cite{MS1}
\bea
w_\pm &=& \left( \frac{r^2 - r_+^2}{r^2 - r_-^2} \right)^\shalf
 e^{2\pi T_\pm u_\pm}, \nn
y &=& \left( \frac{r_+^2 - r_-^2}{r^2 - r_-^2} \right)^\shalf
 e^{\pi ( T_+ u_+ + T_- u_-)},
\label{co1}
\ena
with
\EQ
T_\pm = \frac{r_+ \mp r_-}{2\pi}, \;\;
u_\pm = \phi \pm t.
\label{co11}
\EN
In the region $r>>r_\pm$ in the BTZ coordinates, eqs.~\p{co1} can be
approximated as
\bea
w_\pm = e^{2\pi T_\pm u_\pm}, \;\;
y = \left( \frac{r_+^2 - r_-^2}{r^2} \right)^\shalf
 e^{\pi ( T_+ u_+ + T_- u_-)}.
\label{co2}
\ena

Since the boundary field $\phi_0(w_+,w_-)$ has conformal dimension
$(h_-,h_-)$, it is easy to see that near the boundary $y \to 0(r\to\infty)$,
the bulk field $\phi(r,u_+,u_-)$ in BTZ coordinates behaves as
\bea
\phi (r,u_+,u_-) \sim (2\pi T_+)^{-h_-}(2\pi T_-)^{-h_-}
 (r_+^2 - r_-^2)^{h_-} r^{-2h_-} \phi_0(u_+,u_-).
\label{asy}
\ena
By use of the conformal dimensions of $\phi_0(w_+,w_-)$
and the relations \p{co2}, eq.~\p{sol} in the region $r>>r_\pm$ is cast
into
\EQ
\phi(r,u_+,u_-) = \int du_+' du_-' K_B(r,u_+,u_-;u_+',u_-')\phi_0(u_+',u_-'),
\label{sol1}
\EN
with
\bea
K_B (r,u_+,u_-;u_+',u_-') &=& \frac{2h_+-1}{\pi}
 \left(\frac{r_+^2 - r_-^2}{r^2}\right)^{h_+}
 (2\pi T_+)^{-h_+}(2\pi T_-)^{-h_+} \nn
&& \hs{-35} \times \left\{ \frac{\pi^2 T_+ T_-}{\frac{r_+^2 - r_-^2}{4r^2}
 e^{\pi[T_+ (u_+ -u_+')
 + T_- (u_- -u_-')]} + \sinh \pi T_+( u_+ -u_+') \sinh \pi T_-( u_- -u_-')}
\right\}^{2h_+}.
\label{kb}
\ena
In the derivation of \p{kb}, we keep all the coefficients in the calculation,
and use the normalized bulk-to-boundary Green function.

According to AdS$_3$/CFT correspondence, the relation between string
theory in the bulk and field theory on the boundary is~\cite{W}
\EQ
e^{-S_{eff}(\phi)} = \lan e^{\int_B \phi_0 {\cal O}} \ran_{CFT}.
\label{for}
\EN
Since $\phi$ is the solution to equations of motion, the bulk contribution
to $S_{eff}(\phi)$ is zero, and a boundary term contributes to it:
\bea
S_{eff}= \lim_{r\to \infty} \shalf \int du_+ du_- \sqrt{g} g^{rr}
\phi \pa_r \phi.
\ena
Combining \p{met} and \p{asy}-\p{kb}, we find
\bea
S_{eff} &=& - \frac{h_+(2h_+-1)}{\pi} (r_+^2 - r_-^2)
 (2\pi T_+)^{-1}(2\pi T_-)^{-1} \nn
&& \times \int du_+ du_- du_+' du_-' \phi_0(u_+,u_-)
\left( \frac{\pi T_+}{\sinh \pi T_+ (u_+ -u_+')} \right)^{2h_+} \nn
&& \hs{15} \times
\left( \frac{\pi T_-}{\sinh \pi T_- (u_- -u_-')} \right)^{2h_+}
\phi_0(u_+',u_-').
\label{sol2}
\ena
From \p{for} and \p{sol2}, one has
\bea
G(t,\phi) &=& \lan {\cal O}(u_+,u_-) {\cal O}(0,0) \ran \nn
&=&  \frac{2h_+(2h_+-1)}{\pi}
\left( \frac{\pi T_+}{\sinh \pi T_+ u_+} \right)^{2h_+}
\left( \frac{\pi T_-}{\sinh \pi T_- u_-} \right)^{2h_+},
\label{g1}
\ena
where we have used \p{co11} to simplify the expression.\footnote{
According to ref.~\cite{FMM}, the Ward identities suggest that the factor
$h_+$ in eq.~\p{g1} may be modified to $2 h_+ -1$ due to singular nature
of the two-point correlation functions. This affects the following
expressions, but the results for massless particles remain the same.}
However, due to the periodic identification $\phi \sim \phi + 2 \pi$,
the above expression for $G(t,\phi)$ should be modified by the method of
images as~\cite{K}
\bea
G_T(t,\phi) &=& \lan {\cal O}(u_+,u_-) {\cal O}(0,0) \ran \nn
&& \hs{-15} = \frac{2h_+(2h_+-1)}{\pi}
\sum_{n=-\infty}^{\infty}
\left( \frac{\pi T_+}{\sinh \pi T_+(\phi + t + 2n\pi) } \right)^{2h_+}
\left( \frac{\pi T_-}{\sinh \pi T_-(\phi - t + 2n\pi) } \right)^{2h_+}.
\label{z1}
\ena
The sum over $n \neq 0$ in \p{z1} comes from the twisted sectors of operator
${\cal O}(u_+,u_-)$ in the orbifold procedure $u_\pm \sim u_\pm
+ 2n\pi $ for the BTZ black holes ~\cite{MS1}.
The greybody factors in BTZ black holes are given by~\cite{G,T}
\bea
\sigma_{abs} &=& \frac{\pi}{\omega} \int dt \int_{0}^{2\pi}d\phi e^{ip\cdot x}
 [ G_T(t-i\e,\phi) - G_T(t+i\e, \phi)] \nn
&=& \frac{\pi}{\omega} \int dt \int_{-\infty}^{\infty}d\phi e^{ip\cdot x}
 [ G(t-i\e,\phi) - G(t+i\e, \phi)] \nn
&=& \frac{2h_+(2h_+-1) (2\pi T_+ l)^{2h_+-1}(2\pi T_- l)^{2h_+-1}
 \sinh\left( \frac{\omega}{2 T_H} \right)}{\omega \Gamma^2(2h_+)} \nn
&& \times \left| \Gamma\left(h_+ + i\frac{\omega}{4\pi T_+} \right)
 \Gamma\left(h_+ + i\frac{\omega}{4\pi T_-} \right) \right|^2,
\label{genf}
\ena
where the infinite sum in $G_T(t,\phi)$ has changed the original
integral region $ 0 \leq \phi \leq  2\pi $ into $ {-\infty} \leq \phi \leq
{\infty} $  and the parameter $l$  has been switched on. The Hawking
temperature $T_H$ is defined by
\EQ
\frac{2}{T_H} = \frac{1}{T_+} + \frac{1}{T_-}.
\EN
Here we should point out that in the usual derivation of greybody
factors in 5D and 4D black holes~\cite{MS2,G}, the periodicity 
along the spatial direction $\phi$ is ignored, because it is
assumed that the length of effective string is large compared
to the typical wavelength of the particle, i.e., 2$\pi$ is
much larger than $\frac{1}{T_H}$. However, in our method the same
result can be obtained without the above assumption due to the
good behavior of $G_T(t,\phi)$.

In the $m^2 l^2 \to 0$ limit ($h_+=1$), the decay rate for massless
scalar field is
\bea
\Gamma &=& \frac{\sigma_{abs}(h_+=1)}{e^{\omega/T_H}-1} \nn
&=& \frac{\omega \pi^2 l^2}{(e^{\omega/2T_+}-1)(e^{\omega/2T_-}-1)},
\ena
which is consistent with semiclassical gravity calculations
in~\cite{BSS,LM}. We note that there is a minor difference between \p{genf}
and that obtained in ref.~\cite{LM} with $h_L=h_R=h_+$. In eq.~\p{genf}
there is an extra factor $h_+$, however, when $h_+=1$, both coincide.

The agreement of greybody factors in BTZ black holes obtained from
AdS$_3$/CFT correspondence with those from semiclassical gravity
calculations indicates that the usual effective string theory is nothing
but the boundary 2D CFT of AdS$_3$ space. Let us recall that the two-point
correlation functions in Poincar\'{e} and BTZ coordinates can be written
as
\bea
&& \hs{-10} \lan {\cal O}(w_+,w_-){\cal O}(w_+',w_-')\ran \sim
 \frac{1}{(w_+ - w_+')^{2h_+}(w_- - w_-')^{2h_+}}, \nn
&& \hs{-10} \lan {\cal O}(u_+,u_-){\cal O}(u_+',u_-')\ran \sim
 \frac{2h_+(2h_+-1)}{\pi}
\left[ \frac{\pi T_+}{\sinh\pi T_+ (u_+ - u_+')}\right]^{2h_+}
\left[ \frac{\pi T_-}{\sinh\pi T_- (u_- - u_-')}\right]^{2h_+}
\label{cor}
\ena
In \p{cor}, the Poincar\'{e} coordinates $w_\pm$ are related to BTZ
coordinates $u_\pm$ by an exponential transformation \p{co2} in the
region $r>> r_\pm$. Comparing \p{cor} with eqs.~\p{o}-\p{g}, we find
that ${\bar z}$ and $z$ are not linear in $x^+$ and $x^-$, but rather they
should be related by an exponential transformation. The result obtained
also shows that the initial state of BTZ black holes can be described
in 2D CFT by Poincar\'{e} vacuum state. The operators ${\cal O}_+(w_+)$
and ${\cal O}_-(w_-)$ in Poincar\'{e} coordinates satisfy the OPE in
eq.~\p{ope}. However, the nonlinear coordinate transformation~\p{co2}
introduces a mapping of the original operator ${\cal O}(w_+,w_-)$ in
Poincar\'{e} coordinates to the new one ${\cal O}(u_+,u_-)$ in BTZ
coordinates, and this induces the Bogoliubov transformation on the
operators. The operator ${\cal O}(u_+,u_-)$ see Poincar\'{e} vacuum state
as an excited mixed state; that is, they see the Poincar\'{e} vacuum
state as thermal bath of excitations in BTZ modes~\cite{MS1,K}. The
usual procedure to thermally average the initial state of black holes
(or scalar particles) in the calculation of greybody factors is just
to measure Poincar\'{e} vacuum state by the operator ${\cal O}(u_+,u_-)$
in BTZ coordinates, which was vague in the former treatment of greybody
factors by effective string model in 5D and 4D black holes.

Having understood the greybody factors in BTZ black holes, let us now
discuss the greybody factors in 5D and 4D black holes in the light of
the fact that in the large $N$ limit, the geometries of 5D and 4D black
holes are BTZ $\times S^3 \times M_4$ and BTZ $\times S^2 \times M_5$,
respectively~\cite{MS1,LB,T}. For example, consider near-horizon
AdS$_3$ structure of 5D black holes (``boosted'' D1/D5 configuration)
in the large $N$ limit~\cite{M}. Its metric is~\cite{MS1,LM}
\bea
ds^2 = \frac{r^2}{R^2} (-dt^2 + dx^2) + \frac{r_0^2}{R^2}
 (\cosh \sigma dt + \sinh \sigma dx)^2 + \frac{R^2}{r^2-r_0^2} dr^2
 + R^2 d\Omega_3^2 + \frac{r_1^2}{R^2} \sum_{i=1}^4 dx_i^2,
\label{met2}
\ena
where $r_0$ is the extremality parameter, $r_1, r_5$ and $r_0 \sinh \sigma$
are related to the charges of D1-brane $(Q_1)$, D5-brane $(Q_5)$ and
momentum in 5D black holes, and $R^2 = r_1 r_5$. The $(t,x,r)$ part of
the metric \p{met2} is the metric of BTZ black holes. The
coordinates $(t,x)$ in \p{met2} can be used to construct BTZ coordinates
$u_\pm$, from which the Poincar\'{e} coordinates $w_\pm$ can be introduced,
and 2D CFT lives on the asymptotic boundary $r \to \infty$~\cite{MS1}.

In the background~\p{met2}, two-point correlation function is modified
by a factor $\eta_{5D}$~\cite{FMM,T,OZ}
\bea
\lan {\cal O}(w_+,w_-) {\cal O}(w_+',w_-') \ran
= \eta_{5D} \frac{2h_+(2h_+ -1)}{\pi} \frac{1}
 {(w_+ -w_+')^{2h_+}(w_- -w_-')^{2h_+}},
\ena
where the constant $\eta_{5D}$ can be completely determined from the parent
10D supergravity theory, that is, from the geometry of \p{met2} by the
procedure in refs.~\cite{FMM,MS1,T,OZ}. Following the discussion in BTZ
black holes, it is easy to see that the greybody factors in 5D black holes
is $\sigma^{5D}_{abs} = \eta_{5D} \sigma^{BTZ}_{abs}$. Similarly this
conclusion is also true for 4D black holes. This indicates that the
greybody factors in 5D and 4D black holes have their own origin in BTZ
black holes, and the dynamical information of 5D and 4D black holes are
encoded in BTZ black holes. Thus the boundary dynamics of BTZ black holes,
which is controlled by 2D CFT, looks like hologram and constrains the
essential information of 5D and 4D black holes.

As we have seen, the conformal dimension of operator ${\cal O}(w_+,w_-)$
dominates the greybody factors, and one need not determine the explicit form
of the operator ${\cal O}(u_+,u_-)$. Now let us consider the explicit form
for the $(h,{\bar h})=(1,1)$ operator. For D1/D5 system in type IIB string
theory compactified on $T^4$, the ${\cal N}=(4,4)$ 2D SCFT can be described
by the resolution of the orbifold $(T^4)^{Q_1 Q_5}/S(Q_1 Q_5)$~\cite{SV}.
By AdS/CFT correspondence, the (1,1) operator can be determined from
the symmetries~\cite{DMW2}
\bea
{\cal O}_{ij} = \pa x_A^i {\bar \pa} x_A^j,
\label{op1}
\ena
where $x_A^i$ are the scalar fields in the SCFT under consideration,
$A=1,2,\cdots, Q_1 Q_5$ and $i$ is the vector index of $SO(4)$,
the local Lorentz group of $T^4$. The other possible forms for
$(h,{\bar h})=(1,1)$ operator can be excluded by the symmetries in AdS/CFT
correspondence~\cite{DMW2}. The interaction between ${\cal O}_{ij}$ and
minimal scalars $h_{ij}$ (whose origin is the traceless symmetric
deformations of the 4-torus in type IIB string compactified on $T^4$)
is given by
\EQ
S_{int}= \int d^2 z h_{ij} \pa x_A^i {\bar \pa} x_A^j.
\EN
On the other hand, the (1,1) operator can be introduced by quantizing BTZ
black holes in 3D pure gravity. In ref.~\cite{ES}, it has been shown that
the gauge potentials can be parametrized by
\bea
A_\phi &=& \left( \begin{array}{cc}
a^3(u_+) & e^{-\rho} a^+(u_+) \\
e^{\rho} a^-(u_+) & -a^3(u_+)
\end{array} \right) , \nn
{\tilde A}_\phi &=& - \left( \begin{array}{cc}
{\tilde a}^3(u_-) & e^{-\rho} {\tilde a}^+(u_-) \\
e^{\rho} {\tilde a}^-(u_-) & -{\tilde a}^3(u_-)
\end{array} \right) ,
\ena
and the asymptotic metric of BTZ black holes takes the form
\bea
ds^2 = l^2 d\rho^2 - l^2 e^{2 \rho} a^-(u_+){\tilde a}^+(u_-) du_+ du_-
 + \cdots \; ,
\ena
where the irrelevant subleading terms at large $\rho$ have been omitted.
Then the action~\p{euc} is transformed into
\bea
S_{eff} = \int du_+ du_- {\cal O}_{gravity}(u_+,u_-) \sin(\omega t-n\phi),
\label{act1}
\ena
where
\bea
{\cal O}_{gravity}(u_+,u_-)= a^-(u_+) {\tilde a}^+(u_-),
\label{op2}
\ena
and the classical solution with its asymptotic form
\bea
\phi(\rho,t,\phi) = (1-i e^{-2\rho}) e^{i(\omega_+ t - n_+ \phi)}
 + (1+i e^{-2\rho}) e^{i(\omega_- t - n_- \phi)},
\ena
has been exploited to get \p{act1} with $\omega= \omega_+ - \omega_-$ and
$n= n_+ - n_-$. Comparing \p{op2} with \p{op1}, one is led to the
identification
\bea
\pa x_A {\bar \pa} x_A \leftrightarrow a^-(u_+) {\tilde a}^+(u_-),
\ena
which shows that the contribution to ${\cal O}(u_+,u_-)$ from different
$x_A^i$ is smeared as seen by the operator ${\cal O}_{gravity}(u_+,u_-)$.
Namely 3D pure gravity cannot distinguish between different CFT states
with the same expectation value for the operator ${\cal O}(u_+,u_-)$.
From these observations, one concludes that 3D gravity is a kind of
thermodynamics but gauge theory is statistical mechanics~\cite{M2}.

In the above discussion, we have only considered the greybody factors
induced by massive scalar fields. It is highly nontrivial to check whether
the above identifications for the initial state of black holes in 2D CFT
hold valid also for the spinor field case in the AdS/CFT correspondence.

As we have seen, the $(h,{\bar h})=(1,1)$ operator ${\cal O}(u_+,u_-)$ 
can be easily
obtained by quantizing BTZ black holes in 3D gravity, however,
it is unclear
whether we can get operators ${\cal O}(u_+,u_-)$ of higher
conformal dimensions by quantization of 3D gravity. If not, it is worth
discussing whether and how they can be induced in the context of
six-dimensional supergravity on AdS$_3 \times S^3$, since the
Kaluza-Klein spectrum of 6D supergravity truncated by `stringy exclusion
principle' matches the spectrum of 2D SCFT~\cite{MS1,LB}.

Recently it has been argued that
the isometry group $SL(2,R)$ of quantum gravity on AdS$_2$ can be enlarged
to the full infinite-dimensional $1+1$ conformal group, and the mapping
AdS$_3 \to$ AdS$_2$ has been found~\cite{S}. It would be also interesting
to see whether it is possible to find the origin of greybody factors
in AdS$_2$ context as well.

We hope to return to these issues in near future.

\section*{Acknowledgements}

We would like to thank E. Keski-Vakkuri, D. Kutasov and J.-L. Petersen for
useful discussions. N.O. would also like to thank Niels Bohr Institute for
their hospitality while part of this work was carried out. The work was
supported in part by the Japan Society for the Promotion of Science and
Danish Research Academy.

\newcommand{\NP}[1]{Nucl.\ Phys.\ {\bf #1}}
\newcommand{\AP}[1]{Ann.\ Phys.\ {\bf #1}}
\newcommand{\PL}[1]{Phys.\ Lett.\ {\bf #1}}
\newcommand{\CQG}[1]{Class. Quant. Gravity {\bf #1}}
\newcommand{\NC}[1]{Nuovo Cimento {\bf #1}}
\newcommand{\CMP}[1]{Comm.\ Math.\ Phys.\ {\bf #1}}
\newcommand{\PR}[1]{Phys.\ Rev.\ {\bf #1}}
\newcommand{\PRL}[1]{Phys.\ Rev.\ Lett.\ {\bf #1}}
\newcommand{\PRE}[1]{Phys.\ Rep.\ {\bf #1}}
\newcommand{\PTP}[1]{Prog.\ Theor.\ Phys.\ {\bf #1}}
\newcommand{\PTPS}[1]{Prog.\ Theor.\ Phys.\ Suppl.\ {\bf #1}}
\newcommand{\MPL}[1]{Mod.\ Phys.\ Lett.\ {\bf #1}}
\newcommand{\IJMP}[1]{Int.\ Jour.\ Mod.\ Phys.\ {\bf #1}}
\newcommand{\JP}[1]{Jour.\ Phys.\ {\bf #1}}


\begin{thebibliography}{99}
\bibitem{M} J. Maldacena, preprint, hep-th/9711200.
\bibitem{GKP} S. Gubser, I. Klebanov and A. Polyakov, preprint, hep-th/9802109.
\bibitem{W} E. Witten, preprint, hep-th/9802150.
\bibitem{MS1} J. Maldacena and A. Strominger, preprint, hep-th/9804085.
\bibitem{M1} E. Martinec, preprint, hep-th/9804111.
\bibitem{BBG} K. Behrndt, I. Brunner and I. Gaida, preprint,
hep-th/9804159; hep-th/9806195.
\bibitem{LB} F. Larsen, preprint, hep-th/9805208;
 J. de Boer, preprint, hep-th/9806104.
\bibitem{GKS} A. Giveon, D. Kutasov and N. Seiberg, preprint, hep-th/9806194.
\bibitem{DMW1} A. Dhar, G. Mandal and S. Wadia, \PL{B388} (1996) 51,
 hep-th/9605234;
 S. Das and S.D. Mathur, \NP{B478} (1996) 561, hep-th/9606185; \NP{B482} (1996)
 153, hep-th/9607149. 
\bibitem{MS2} J. Maldacena and A. Strominger, \PR{D55} (1997) 861,
 hep-th/9609026; \PR{D56} (1997) 4975, hep-th/9702015.
\bibitem{CGK} C.G. Callan, S. Gubser, I. Klebanov and A.A. Tseytlin, \NP{B489}
 (1997) 65, hep-th/9610172;
 I. Klebanov and M. Krasnitz, \PR{D55} (1997) 3250, hep-th/9612051; \PR{D56}
 (1997) 2173, hep-th/9703216;
 S. Gubser, \PR{D56} (1997) 4984, hep-th/9704195; M. Cvetic and
F. Larsen, preprints, hep-th/9706071; hep-th/9712118.
\bibitem{G} S. Gubser, \PR{D56} (1997) 7854, hep-th/9706100.
\bibitem{T} E. Teo, preprint, hep-th/9805014.
\bibitem{DMW2} J. David, G. Mandal and S. Wadia, preprint, hep-th/9808168.
\bibitem{BTZ} M. Ba\~{n}ados, C. Teitelboim and J. Zanelli, \PRL{69} (1992)
 1849, hep-th/9204099.
\bibitem{SSH} K. Sfetsos and K. Skenderis, \NP{B517} (1998) 179,
 hep-th/9711138; S. Hyun, preprint, hep-th/9704005.
\bibitem{BSS} D. Birmingham, I. Sachs and S. Sen, \PL{B413} (1997) 281,
 hep-th/9707188.
\bibitem{LM} H.W. Lee and Y.S. Myung, preprints, hep-th/9808002;
 hep-th/9804095.
\bibitem{HSM} M. Henningson and K. Sfetsos, preprint, hep-th/9803251;
 W. Muck and K. Viswanathan, preprints, hep-th/9804035; hep-th/9805145;
 H. Liu and A.A. Tseytlin, preprints, hep-th/9804083; hep-th/9807097;
 T. Banks and M. Green, preprint, hep-th/9804170;
 G. Chalmers, H. Nastase, K. Schalm and R. Siebelink, preprint, hep-th/9805105;
 A. Chezelbach, K. Kaviani, S. Parvizi and A. Fatollahi, preprint,
 hep-th/9805162;
 S. Lee, S. Minwalla, M. Rangamani and N. Seiberg, preprint, hep-th/9806074;
 G. Arutyunov and S. Frolov, preprint, hep-th/9806216;
 E. D'Hoker, D.Z. Freedman and W. Skiba, preprint, hep-th/9807098;
 S. Corley, preprint, hep-th/9808184; A. Volvovich, preprint, hep-th/9809009.
\bibitem{FMM} D.Z. Freedman, S.D. Mathur, A. Matusis and L. Rastelli,
 preprints, hep-th/9804058; hep-th/9808006.
\bibitem{BKL} V. Balasubramanian, P. Kraus and A. Lawrence, preprint,
 hep-th/9805171;
 V. Balasubramanian, P. Kraus, A. Lawrence and S. Trivedi, preprint,
 hep-th/9808017.
\bibitem{K} E. Keski-Vakkuri, preprint, hep-th/9808037.
\bibitem{OZ} N. Ohta and J.-G. Zhou, \NP{B522} (1998) 125, hep-th/9801023.
\bibitem{M2} E. Martinec, preprint, hep-th/9809021; M. Li, E. Martinec and
 V. Sahakian, preprint, hep-th/9809061.
\bibitem{SV} A. Strominger and C. Vafa, \PL{B379} (1996) 99, hep-th/9601029;
 C. Vafa, \NP{B463} (1996) 435, hep-th/9512078.
\bibitem{ES} R. Emparan and I. Sachs, preprint, hep-th/9806122.
\bibitem{S} A. Strominger, preprint, hep-th/9809027.
\end{thebibliography}
\end{document}